\documentclass[useAMS,usenatbib]{mn2e}

\usepackage{graphicx}
\usepackage{pifont}

\title[Fractal analysis of young clusters]{Statistical fractal analysis of 25 young star clusters}
\author[Gregorio-Hetem et al.]{J. Gregorio-Hetem$^{1}$\thanks{E-mail:
jane.gregorio.hetem@iag.usp.br}, A. Hetem$^{2}$, T. Santos-Silva$^{1}$, 
B. Fernandes$^{1}$\\
$^{1}$Universidade de S{\~a}o Paulo, IAG, Rua do Mat{\~a}o 1226, 05508-900 S{\~a}o Paulo, Brazil\\
$^{2}$UFABC Universidade Federal do ABC, CECS, Av. dos Estados, 5001, 09210-580 Santo Andr{\'e}, SP, Brazil} 

\begin{document}

\date{Accepted Received }

\pagerange{\pageref{firstpage}--\pageref{lastpage}} \pubyear{2002}

\maketitle

\label{firstpage}

\begin{abstract}
  A large sample of young stellar groups is analysed aiming to investigate 
 their clustering properties  and dynamical evolution. 
 A comparison of the $\mathcal{Q}$ statistical parameter, measured for the clusters, 
 with the fractal dimension estimated for the  projected clouds shows that 52\%
 of the sample has substructures and tends to follow the  theoretically expected relation between 
 clusters and clouds, according to calculations for artificial distribution of points. 
The fractal statistics was also compared to structural parameters revealing that
clusters having radial density profile show a trend of parameter $\overline {s}$ 
increasing with mean surface stellar density.
The core radius of the sample, as a function of age, follows a distribution 
similar to that observed in stellar groups of Milky Way and other galaxies. They also have 
dynamical age, indicated by their crossing time, that is similar to unbound associations.
The statistical analysis allowed us to separate the sample into two groups
showing different clustering characteristics. However, they have the same dynamical evolution, 
since the whole sample has been revealed as expanding objects, for which the substructures seem
to have not been erased. These results are in agreement
with  simulations that adopt low surface densities and models under supervirial conditions.
\end{abstract}


\section{Introduction}
\label{sect1}

It is widely accepted that
the physical conditions for the formation and evolution of stellar clusters can be 
inferred on basis of embedded clusters structure, which depends on the 
distribution of dense gas of their original cloud. 
Numerous clusters embedded in clouds are found in the Galaxy. \citet{ladas}
suggested that most of them shall lose their dynamical equilibrium, been 
dissolved in a few tens of Myr. Based on the surface density profile of
the distribution of members, they proposed two types of cluster structure: 
centrally concentrated and hierarchical. 

\citet{elmfalg96} demonstrated the fractal origin of the relations of
size and mass distribution in the interstellar gas, determined from clouds surveys 
in the literature. More recently, \citet{girichidis} used hydrodynamic simulations of
formation of filaments and protostars, in different initial configurations, aiming to
evaluate the statistical properties of embedded clusters in the early star-forming
phases. They found that more substructures are formed by flat density 
profiles and compressive modes, instead of to produce centrally concentrated profiles.

Classical open clusters are characterized by a centrally concentrated surface 
distribution, whose radial profile is smooth and can be reproduced by a 
simple power law function. On the other hand, hierarchical type clusters show multiple 
peaks, or substructures in their surface density distribution, like NGC~1333
\citep{lada96}, for instance. More recently, \citet{guter08} discussed the structure of this
cluster based on {\it Spitzer} observations of 137 members, whose surface density
map shows an elongated distribution, not centrally concentrated, suggesting a
low velocity dispersion and that NGC~1333 is not in virial equilibrium.

In spite of the large number of different clusters 
found in the Galaxy, their relative occurrence remains to be verified,
requiring the investigation of clustering properties in a as large as possible
sample of studied objects.

For instance, \citet{feigelson11} studied the structure of a huge sample of young
stars detected by ray-X observations in the {\it Chandra} Carina Complex Project (CCCP).
They found three different types of clustering properties in their sample that includes 
previously known clusters and some additional smaller groups identified by CCCP. 
The classical, centrally concentrated clusters, present an unimodal stellar distribution. 
Objects showing several substructures, distributed in a circular boundary, are classified 
as multimodal clumpy structure, which is suggested to be an unequilibrated stage of cluster 
formation. The third type of cluster contains sparse compact groups, without any concentration 
that is possibly due to triggered star formation.

The effects of ionizing fronts on triggered star formation have been evaluated 
using numerical simulations by \citet{walch13}, for instance.

Considering the rapid changes on the spatial distribution of forming stars shown
by these simulations, it 
is expected a strong evolution of the clustering structure, that can be
evaluated by the statistical parameter $\mathcal{Q}$ \citep{cart04}. 
Indeed, a rapid change of $\mathcal{Q}$ is noted in the simulations of ionization-induced star formation 
by \citet{dale12,dale13}. 
However, smooth evolution is also found in the simulations with- and without 
feedback of ionizing source, depending on the initial conditions \citep{pd, pd14}.

Based on fractal statistics, in a previous work we discussed the formation 
and evolution conditions of four young clusters \citep[hereafter FGH12]{bf12}. The preliminary
results revealed that one object of the sample has radial stellar density profile  that is
not similar to the fractal characteristics of the nearby cloud, in a contrary way of
the other three studied clusters.

Aiming to validate, refine and enlarge our previous
investigation of fractal statistics and its possible 
relation to the  initial conditions of cluster formation, in the present work
the analysis is extended to a sample of 21 open star clusters, selected according to
their youth and intermediate distances. The clusters
were characterized by \citet[hereafter SG12]{ss12} by using near-infrared data
to derive their structural parameters based on 
stellar density maps, similar to the study developed by \citet{bonatto09},
for instance.

The dependence of fractal statistics on age and crossing time 
was also analysed, aiming to investigate the dynamical evolution that is 
expected for the star distribution in young clusters, which has been extensively 
modelled by theoretical simulations (e.g. \citet{cart04}; \citet{goodw}; \citet{pd}; 
\citet{gieles11}; \citet{pd14}; \citet{pa14}; among others). 
The results were also compared to the observational studies by 
\citet{schkl}, \citet{caballero}, \citet{sa09}, \citet[hereafter PZMG10]{pzmg10}  
and \citet{wright}, for instance.

The paper is organized as follows. Section~\ref{sect2} is dedicated to present the characteristics 
of the selected clusters. Section~\ref{sect3} summarizes the adopted methodology, 
describing the calculations of fractal dimension of clouds, clustering statistics
of surface stellar density, and error estimation by using the bootstrap method.
Section~\ref{sect4} presents the results from the statistical analysis, 
and in Sect.~\ref{sect5} the  dynamical evolution of the sample is discussed.
Finally, the main conclusions are summarized in Sect.~\ref{sect6}.

\section{Selected star clusters}
\label{sect2}

All the selected young clusters have similar angular sizes ($R < 20$ arcmin),
and intermediate distances, which is $d \leqslant$ 2 kpc for most of the objects.
Since the observed E(B-V) is low, varying from 0.1 to 0.7, we assume
that none of them can be considered an embedded cluster.

Table~\ref{tab1} lists the 25 young star clusters analysed in the present work, giving the 
information available in the literature.
For comparison purposes, the four clusters previously studied by FGH12 
have been included in the sample.
The other 21 clusters have structural and fundamental parameters determined by SG12. 

\subsection{Observed radial density profile}

SG12 used surface stellar density to derive the structural parameters of the clusters. 
Membership probabilities, evaluated according to proper movement \citep{dias02},
were used to identify the very-likely members. SG12 also adopted
color-magnitude filters that enable to separate field-stars from 
candidate members, which were also included in the total number of objects
($N_T$) used to determine the stellar density maps. 

Table~\ref{tab1} reproduces the structural parameters obtained by SG12 by 
fitting the observed radial density profile, based on the empirical model from \citet{king}:
$\sigma(r)=\sigma_{bg} + {\sigma_{0}}/({1+(r/r_{c})^{2}})$,
where $\sigma(r)$ is the stellar surface density (stars/arcmin${^2}$), 
 $ \sigma_{bg} $ is the average density measured in a reference region (background), 
$\sigma_{0}$ and $r_{c}$ are respectively the density and the radius of the cluster core.
The cluster radius ($R$) is given by the point where $\sigma(r) = \sigma_{bg}$. The adopted errors 
are the uncertainties provided by the fitting method. 
In Sect.~\ref{sect4} these parameters are 
compared to the statistical fractal parameters obtained in the present work.

\begin{table}
{\scriptsize
\caption{Sample studied in the present work and informations from literature.}
\label{tab1}
\begin{center} 
\begin{tabular}{l c c c c c c}
\hline
Cluster	&	$N_T$	&	Age	&	$d$	&	$n$	&	$R$	&	$r_c$ 	\\
	&		&	(Myr)	&	(pc)	&	(pc$^{-2}$)	&	(pc)	&	(pc)	\\
	\hline	
Collinder~205	&	174	&	5	&	1800	&	4.8$\pm$1.9	&	3.4$\pm$1.2	&	0.30$\pm$0.03	\\
Hogg~10	&	88	&	3	&	2200	&	7.6$\pm$1.7	&	1.9$\pm$0.4	&	0.46$\pm$0.28	\\
Hogg~22	&	97	&	3	&	1700	&	9.0$\pm$3.8	&	1.9$\pm$0.5	&	1.9$\pm$1.1	\\
Lynga~14	&	68	&	3	&	950	&	16.7$\pm$7.0	&	1.1$\pm$0.4	&	0.22$\pm$0.04	\\
Markarian~38	&	57	&	6	&	1500	&	6.8$\pm$2.3	&	1.6$\pm$0.5	&	0.13$\pm$0.03	\\
NGC~2302	&	70	&	3	&	1700	&	4.0$\pm$1.4	&	2.3$\pm$0.5	&	0.54$\pm$0.10	\\
NGC~2362	&	124	&	4	&	1480	&	3.3$\pm$1.0	&	1.9$\pm$0.4	&	0.26$\pm$0.05	\\
NGC~2367	&	60	&	3	&	2200	&	2.3$\pm$0.6	&	2.9$\pm$0.7	&	0.34$\pm$0.11	\\
NGC~2645	&	104	&	5	&	1800	&	7.5$\pm$2.1	&	2.1$\pm$0.5	&	0.22$\pm$0.04	\\
NGC~2659	&	215	&	5	&	2000	&	5.2$\pm$0.9	&	3.6$\pm$0.5	&	1.93$\pm$0.37	\\
NGC~3572	&	149	&	3	&	1900	&	7.7$\pm$1.9	&	2.5$\pm$0.5	&	0.16$\pm$0.03	\\
NGC~3590	&	79	&	3	&	1680	&	11.7$\pm$4.5	&	1.5$\pm$0.4	&	0.29$\pm$0.06	\\
NGC~5606	&	98	&	3	&	2200	&	4.8$\pm$0.9	&	2.6$\pm$0.4	&	0.51$\pm$0.13	\\
NGC~6178	&	106	&	5	&	1430	&	12.2$\pm$2.7	&	1.7$\pm$0.4	&	0.30$\pm$0.12	\\
NGC~6604	&	90	&	6	&	1600	&	4.4$\pm$1.3	&	2.6$\pm$0.7	&	0.50$\pm$0.11	\\
NGC~6613	&	133	&	5	&	1550	&	4.9$\pm$1.2	&	2.9$\pm$0.7	&	0.12$\pm$0.01	\\
Ruprecht~79	&	174	&	5	&	2700	&	3.6$\pm$0.5	&	3.9$\pm$0.5	&	2.20$\pm$0.80	\\
Stock~13	&	65	&	4	&	2000	&	3.8$\pm$0.9	&	2.3$\pm$0.3	&	0.11$\pm$0.05	\\
Stock~16	&	139	&	7	&	2000	&	5.8$\pm$1.6	&	2.8$\pm$0.7	&	1.20$\pm$0.60	\\
Trumpler~18	&	164	&	5	&	2850	&	2.5$\pm$0.4	&	4.6$\pm$0.8	&	0.38$\pm$0.10	\\
Trumpler~28	&	73	&	2	&	1050	&	12.3$\pm$4.5	&	1.4$\pm$0.5	&	0.84$\pm$0.16	\\
\hline
Berkeley~86	&	85	&	5	&	1585	&	3.4	&	2.8	&		\\
NGC~2244	&	295	&	3	&	1660	&	1.9	&	7.0	&		\\
NGC~2264	&	292	&	3	&	760		&	5.0	&	4.3	&		\\
NGC~6530	&	62	&	3	&	1300	&	2.9	&	2.6	&		\\
\hline													

\end{tabular}
\end{center}}
{
\footnotesize
Note: Data for 21 clusters are given by SG12, and for the last four clusters are from FGH12 and 
references therein.
}
\end{table}

\subsection{Mass and age}

By comparing the cluster members with pre-MS \citep{siess} and main sequence \citep{girardi} models in 
the unreddened near-IR colour-magnitude diagrams, SG12 estimated fundamental 
parameters like age and mass. The mean ages are $\sim 5$ Myr,
but some of the clusters have a bimodal age distribution, showing a second peak around
12 Myr (Hogg~22, Lynga~14, NGC~2302, NGC~2645, NGC~5606, and NGC~5606).

The observed mass distribution of each cluster was fitted with the mass function 
$\phi(m) \propto m^{-(1+\chi)}$ from \citet{kroupa}.
 The $\chi$ values determined by the fitting method indicate that 
33\% of the sample have low power-law index, comparable to the slope $\chi = 0.3 \pm 0.5$
assumed by \citet{kroupa} for $M < 0.5 \rm{M}_{\odot}$. The other 67\% of the clusters have 
deeper slopes $\chi \sim 0.8$ to 1.5 that are, within the errors, 
roughly similar to the Salpeter's IMF ($\chi = 1.35$), 
which is the index that \citet{kroupa} suggested for $M> 1 \rm{M}_{\odot}$. 
This result is also in agreement with the 
mass distribution with $\chi \sim 1.0$  that has been found for the initial
cluster mass function of diverse clusters \citep[e.g.][]{elmegreen10,oey}. 

The mean mass of the cluster members is $>1~\rm{M}_{\odot}$, indicating an 
incompleteness of low-mass stars, probably due to the 2MASS detection limit 
that constrains the presence of faint sources in the sample.

Considering the total mass, it is interesting to note that our sample
corresponds to an intermediate range between the embedded clusters studied by \citet{ladas}
and the massive ``leak" clusters having $R >$ 10 pc in the sample studied by \citet{pfal11}. 
All of them follow the same mass-radius relation given by $M \sim 118 R^{1.3}$.
 SG12 (see their Fig.~4c) compare the sample with the distribution of clusters analysed by
\citet{pfal11} who proposed that these objects represent the ending of a time sequence, which
mass-radius dependence is $M \sim 359 R^{1.7}$. However, this power law fitting is significantly 
different of the distribution mentioned above, found by SG12 for our sample, as well as for the 
embedded clusters from \citet{ladas} and also for clusters showing $R <$ 10 pc 
in the sample analysed by \citet{pfal11}.

The results from SG12 favour the interpretation given by \citet{adams06} and \citet{adams10} suggesting that
differences on mass-radius relations are more likely due to different formation conditions
instead of a time sequence ending in the massive ``leak" clusters proposed by \citet{pfal11}.

A similar conclusion was achieved  by FG12 from the estimated volumetric density, which is correlated
with cluster radius by $\rho = 28 R^{-1.7}$, where $\rho$ is given in M$_{\odot}$/pc$^3$,
which stands for our sample as well as for embedded clusters from \citet{ladas}, being 
quite similar to other studies, like $\rho \propto R^{-1.92}$ found by \cite{camargo10}, 
for instance.

\begin{figure*}
\begin{center}
\includegraphics[width=5.2cm,angle=270]{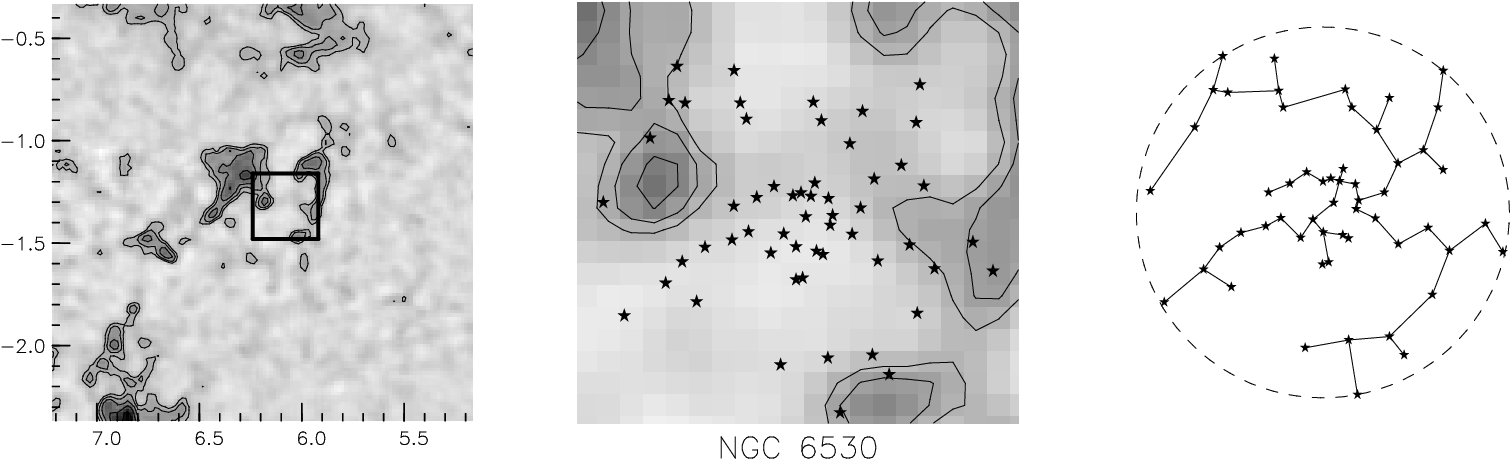}
\vskip 0.3cm
\includegraphics[width=5.2cm,angle=270]{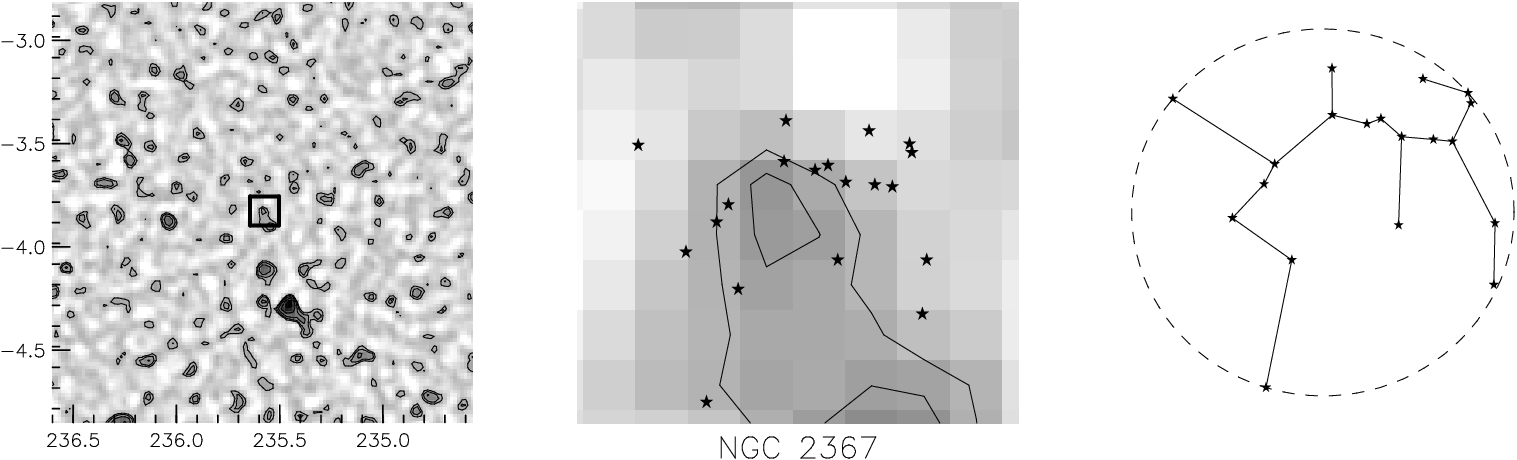}
\vskip 0.3cm
\includegraphics[width=5.2cm,angle=270]{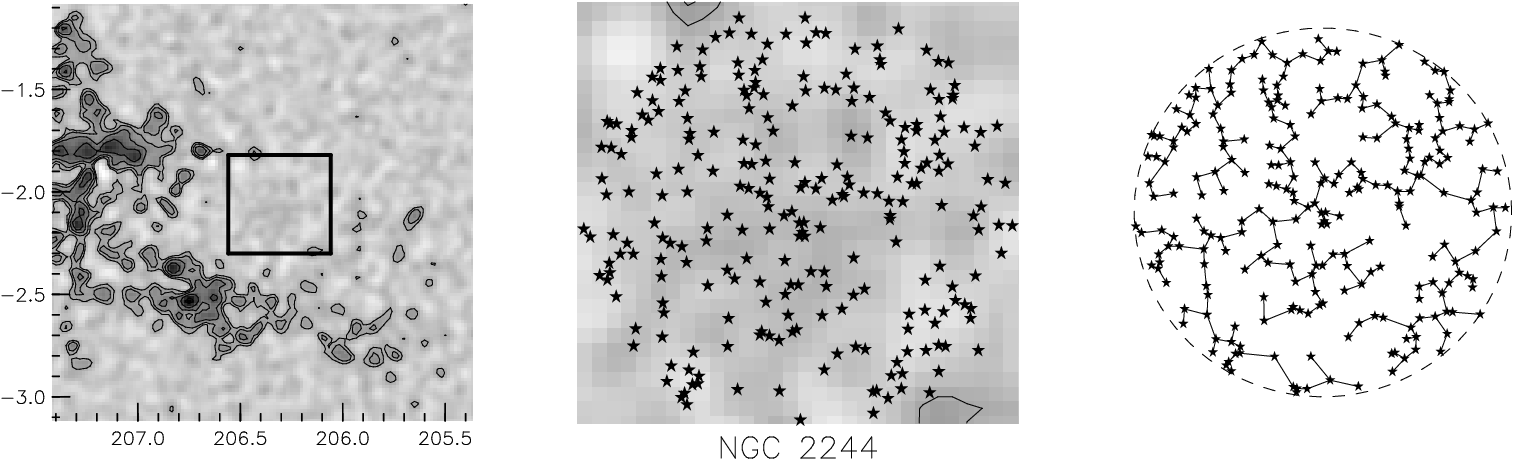}
\vskip 0.3cm
\includegraphics[width=5.2cm,angle=270]{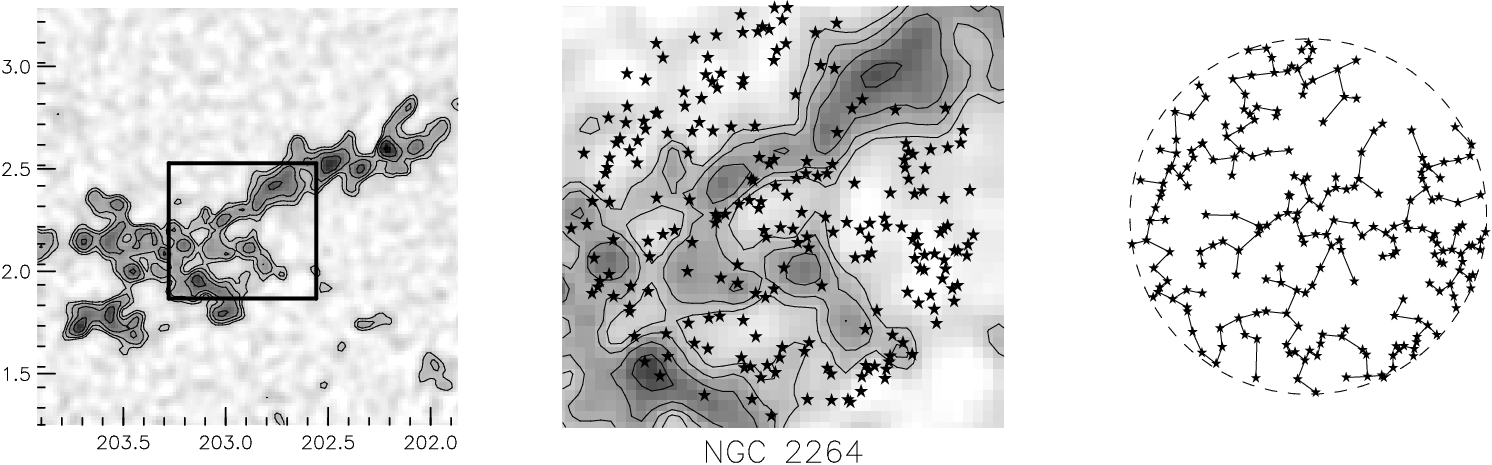}
\caption{ 
{\it Left:} Visual extinction ($A_V$) map (galactic coordinates) of nearby clouds of four objects of the sample.
 Contours represent $A_V$ levels with S/N $>$ 10. 
The location of the star clusters is indicated by the central rectangle. 
{\it Middle:} A zoom of the $A_V$ map showing the spatial distribution of the cluster members. 
{\it Right:} Minimal spanning tree and smallest circle (dashed line) that contains all members of the 
cluster. Upper panels show  clusters with $\mathcal{Q} >$ 0.8 (NGC~6530 and NGC~2367), 
while small scale substructures (NGC~2244 and NGC~2264) are seen in lower panels.}
\label{fig:av}
\end{center}
\end{figure*}


\section{Methodology}
\label{sect3}

FGH12 performed fractal analysis of the clusters Berkeley~86, 
NGC~2244, NGC~2264 and NGC~6530. By comparing with the statistic derived 
from artificial data simulating stars distributions and parental clouds, the 
clustering parameters of three of the analysed clusters are compatible with
the fractal dimension of the observed clouds.
NGC~6530 is the only cluster showing a central concentration of stars that is 
different of the substructures found for the nearby cloud. In the present work 
our previous analysis is extended to a larger sample, 
following the same methodology adopted in FGH12, which is summarized in this section. 


\subsection{Fractal dimension of the nearby clouds}
\label{sect31}

Visual extinction ($A_V$) maps were used to perform statistical analysis of the clouds. Aiming
to investigate the interstellar matter distribution in the direction of the clusters,
the $A_V$ maps were extracted from the Dark Clouds Catalogue presented by \citet{dobashi05}. 

For each cluster we searched for $2^{\rm o}\times 2^{\rm o}$ images containing 
the $A_V$ levels obtained from 2MASS data, with spacial resolution of $\sim 1'$ per pixel. 
A threshold of S/N = 10 was chosen to define the minimum level of $A_V$ contour, 
avoiding low density regions in the calculations. Therefore, $A_V$ maps are useful to
indicate the presence of dense clouds nearby the cluster.

Figure~\ref{fig:av} shows some examples of $A_V$ maps and corresponding spatial distribution of 
 cluster members, which were used in the fractal analysis discussed in Sect.~\ref{sect32}.

Following \citet{hetem93}, the fractal dimension $D_2$ of contour levels is measured
by using the perimeter-area relation of the clouds $P=({A}^{1/2}/{K})^{D_2}$, 
where $P$ is the perimeter of a given $A_V$ contour level and $A$ is the area inside it,
and $K$ is a parameter connected to the geometry (shape) of the objects under study.
The uncertainties on the estimates of $D_2$ and $K$ of projected images 
of clouds were obtained by applying standard least squares methods \citep{press}.

Since this method uses bi-dimensional projected maps, the comparison of $D_2$ with
the values of fractal dimension from literature ($D_3$), determined from three-dimensional 
distributions, requires a $3{\rm D}\times 2{\rm D}$ transforming function ($D_3={\mathcal{F}}(D_2)$).
Following \citet{falc90}, we adopt the relation $D_3 = D_2 + 1$, which is a good approximation
for $D_3 < 2.5$ \citep{cart06}. However, the relationship between fractal dimension of a cloud projection 
and three dimensional calculations may depend on a more complex inference of $D_3={\mathcal{F}}(D_2)$ 
that is not clearly established \citep{sap05}.


\subsection{Clustering statistics}
\label{sect32}

The parameter $\mathcal{Q}$, which carries information about the cluster fractal 
structure, is a dimensionless quantity given by 
$\mathcal{Q} = \overline{m} / \overline {s}$, 
where $\overline{m}$ and $\overline{s}$ are statistical parameters that depend 
on the geometric distribution of data points  \citep{cart04}.

The parameter $\overline{m}$ is the normalized mean edge length, 
related to the surface density of cluster members projected position, defined by:

\noindent $$\overline{m}=\frac{1}{(A_{N} N)^{1/2}} \sum_{i=1}^{N-1}{m_i}$$

\noindent where $N$ is the total number of considered points, 
$m_i$ is the length of edge $i$ in the minimum spanning tree (described below) 
and $A_ {N} $ is the area of the smallest circle that contains all 
points projected on the plane of the cluster. Each smallest circle was determined by
adopting the algorithm proposed by \citet{megiddo}. 

The minimal spanning tree is defined as the unique network of straight lines 
that can connect a set of points without closed loops, 
such that the sum of all the lengths of these lines (or edges)
is the minimal possible. Figure~\ref{fig:av} gives examples of smallest circle 
and respective spanning tree, which were constructed using
the method described by \citet{gower} in order to obtain $\overline{m}$
for each cluster. 

The value of $\overline{s}$ represents the mean separation of the points, 
and is given by

\noindent $$\overline{s} = \frac{2}{N(N-1)R_N} \sum_{i=1}^{N-1}{} 
\sum_{j=1+i}^{N} \vert \overrightarrow{r}_i - \overrightarrow{r}_j \vert$$

\noindent where $r_i$ is the vector position of point $i$, and $R_N$ 
is the radius of the smallest circle that contains all points.

Studies of the hierarchical structure in young clusters have 
used the $\mathcal{Q}$ parameter to distinguish fragmented from smooth 
distributions \citep[e.g.][]{elmegreen10}. 
Comparing the cluster physical structure with the geometry of its possible
parental cloud may tell us about the original gas
distribution of the star forming region and how the particular cloud may have evolved 
during the cluster formation.

Using artificial distributions of points, \citet{lwc11}
adopted the technique proposed by \citet{cart04} to perform  a statistical analysis 
that gives inferences on  the fractal dimension measured on grey-scale images generated 
by models of stars clusters and clouds. They used data sets varying 
from 64 to 65536 points, which give a parameters space of $\mathcal{Q}$ as a function
of radial profiles (index $\alpha$) and fractal dimension ($D_3$) that are similar to, but larger
than the calculations performed by \citet{cart04} and \citet{sa09}, for instance. 
In Sect.~\ref{sect42} the results are compared with these theoretical models.


\subsection{Standard deviation estimated with bootstrapping}

The bootstrap method was adopted in order to estimate the standard deviation of the 
statistical parameters. This numerical method, introduced by \citet{efron}, 
is independent of model or calculations; is not based on asymptotic results; 
and is simple enough to be automatized as an algorithm. 
The method creates a set of simulated data by considering each parameter of the model 
as varying in a chosen confidence level according to a given distribution. 
Then, this artificial set is treated as a real sample 
(as those obtained by observations) 
and the standard deviation is calculated from these data.

In the present work, the bootstrapping algorithm was implemented as proposed by \citet{press}. 
The algorithm generates synthetic distributions of cluster members based on the original 
positions of the stars. A number of stars is moved from their original positions, 
being replaced by random points within a box, around the star, with side equals 
to the cluster radius. 
Then, the parameters $\overline{m}$, $\overline{s}$ and $\mathcal{Q}$ are calculated for 
each synthetic distribution. By adopting a fraction $f = 1/e \sim 37\%$ of 
replaced stars in a symmetric normal distribution, it was possible to derive 
$1\sigma$ deviation for $\mathcal{Q}$, $\overline{m}$ and $\overline{s}$.


\begin{figure}
\begin{center}
\includegraphics[width=6cm,angle=270]{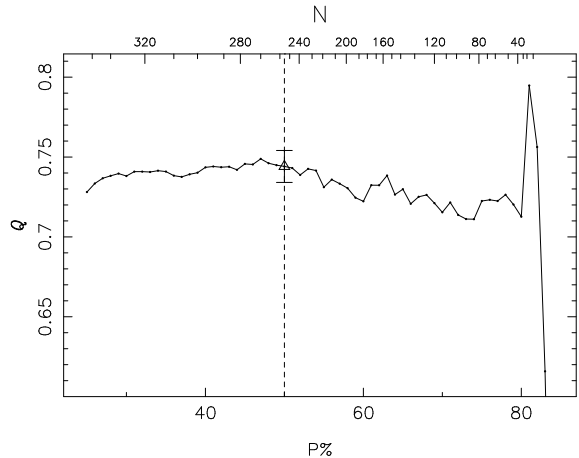}
\caption{Distribution of $\mathcal{Q}$ as a function of number of stars (top axis) and membership
probability (bottom axis) calculated for members and candidate members of NGC~2244. 
The limiting values adopted in
the present work, $N=256$ and $P = 50\%$, are shown by a dashed line.}
\label{fig:qnp}
\end{center}
\end{figure}



\section{Analysis of fractal statistics}
\label{sect4}


\subsection{Membership dependence}
\label{sect41}

In spite of the whole set of stars (candidates and members) had been used
by SG12 to determine the cluster structural parameters, in the present work we preferred to 
constrain the statistical analysis to confirmed members only. 
Based on the probability membership ($P\%$), the stars were separated in two sub-samples: the very 
likely members of the cluster have $P \geq 50\%$ (denoted by $P_m$), while the candidate 
members (called $P_c$) have $P < 50\%$ .

Aiming to evaluate the effect on the fractal statistics by considering 
different numbers of stars and to explore the validity in choosing $P_m$ to 
represent the very likely members, Fig.~\ref{fig:qnp} shows the variation of 
$\mathcal{Q}$ as a function of $N$ for the cluster NGC~2244. 
The adopted result $\mathcal{Q} = 0.74$ obtained for $N=256$ 
members is plotted with error bars. It can be seen that this result stands as a representative value
for almost the whole data set, with a smooth decreasing for low numbers of stars. A significant 
change in $\mathcal{Q}$ occurs only for $N=14$ members with $P>80\%$, but the main conclusion
about the clustering structure of NGC~2244 remains the same.

Figure~\ref{fig:qnp} also displays the variation of $\mathcal{Q}$ as a function of the membership probability. 
This distribution shows that the choice of $P_m$ is good enough
to avoid large errors due to a small data set (low $N$), also excluding possible 
field-stars contamination (low $P\%$).

\begin{figure}
\begin{center}
\includegraphics[width=6cm,angle=270]{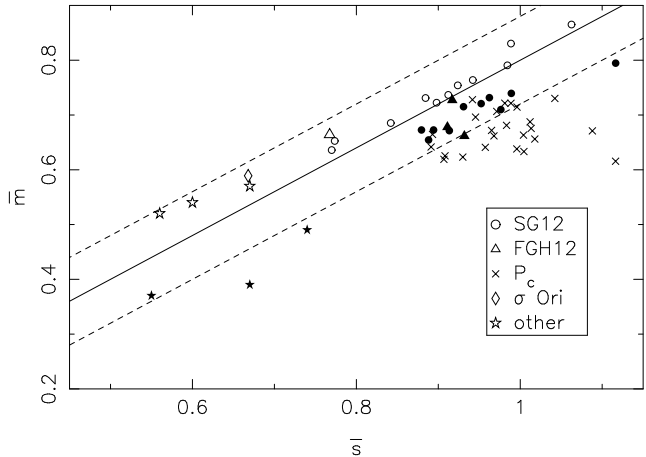}
\caption{Trend of $\overline{m}$ increasing with $\overline{s}$
shown by the very likely members of the sample (SG12 and FGH12).
Candidate members ($P_c$) do not follow this trend that is
illustrated by $\mathcal{Q} = 0.8$ (full line) and
10\% deviation (dashed lines). Data for other clusters from literature 
are also plotted for comparison (see Sect.~\ref{sect41}). 
Filled symbols represent $\mathcal{Q} < 0.8$. 
}
\label{fig:ms}
\end{center}
\end{figure}

For each set of stars ($P_m$ and $P_c$ sub-samples), we performed the calculation of the parameters 
$\overline{m}$ and $\overline{s}$, which are displayed in Fig.~\ref{fig:ms}. It can be seen that the distribution 
of $P_m$ members in this plot tend to follow the expected correlation between $\overline{m}$ 
and $\overline{s}$, while the candidate members  do not follow the same distribution, probably
due to field-stars contamination.

 For comparison with literature, other objects are also included in Fig.~\ref{fig:ms}, like 
$\rho$ Oph, Taurus, Chamaeleon, IC~348, Serpens \citep{schkl}, and IC~2391 which was analysed by 
\citet{cart04} along with the first four clusters mentioned above.
In spite of the agreement on the $\mathcal{Q}$ values between both works, their results
on $\overline{m}$ and $\overline{s}$ differ in almost 20\%, excepting Taurus for which the
$\overline{m}$ values are 50\% discrepant.

Since the cluster $\sigma$ Ori \citep{caballero} has available parameters  that
can be used in several of our comparative plots, a different symbol is used distinguishing it
of other clusters from literature.

\subsection{Cluster structure compared with projected clouds}
\label{sect42}

Table~\ref{tab2} gives the statistical parameters, along with the number of studied stars 
in each cluster, and an estimation of galactic coordinates ({\it l,b}).

Aiming to derive the size of the area occupied by the $P_m$ members, we determined the
mean ({\it l,b}) that give the center of the stellar spatial distribution. The
respective dispersion is used to estimate the observed sizes,
 which show a circular symmetry ($\Delta l \sim \Delta b$) suggesting that our 
objects are not elongated.

\begin{figure}
\begin{center}
\includegraphics[width=6cm,angle=270]{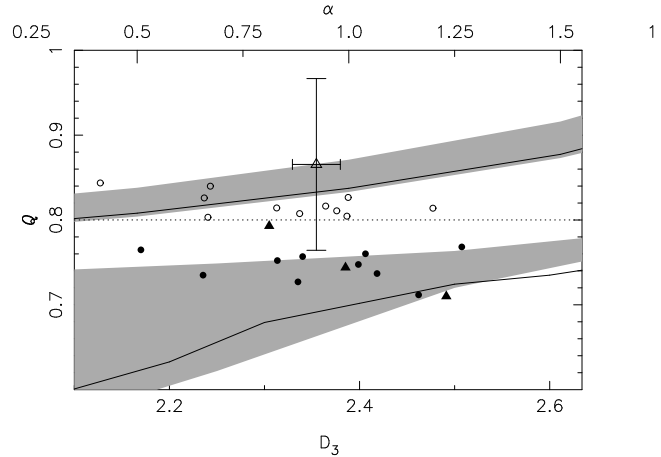}
\caption{Parameter $\mathcal{Q}$, estimated from the spatial distribution of cluster members,
as a function of fractal dimension ($D_3$) of projected clouds. Typical error bars 
are shown for NGC~6530. For comparison, clustering statistics obtained for artificial 
data (shaded areas) and other theoretical results (full lines) are also shown.
\label{fig:QxDf}
}
\end{center}
\end{figure}

In Fig.~\ref{fig:QxDf} the parameter $\mathcal{Q}$, obtained from the $P_m$ spatial distribution, 
is compared with the fractal dimension of the nearby clouds (discussed in Sect.~\ref{sect31}).
The locus of  statistical parameters calculated for artificial data points \citep{lwc11} 
is shown as a shaded area. 
 Above $\mathcal{Q}$ = 0.8 (dotted line) the shaded area shows the dependence on
radial profiles (index $\alpha$), while the variation of fractal 
dimension ($D_3$) is found below this line. 
Since the index $\alpha$ was not estimated for our objects, 
the upper region of Fig.~\ref{fig:QxDf} is displayed 
only for comparison with other theoretical results \citep{sa09}(full lines).
They are also comparable to simulations for cluster types F2.0 to F3.0 and 3D0 to 3D2.9
from \citet{cart04}, which respectively correspond to fractal and 
radial distributions.

 It can be noted that 43\% of the sample is found around
the bottom shaded area of Fig.~\ref{fig:QxDf}, meaning that 
substructures observed for these clusters ($\mathcal{Q} <$ 0.8) coincide with the
fractal characteristics of the projected clouds ($D_3 < 2.5)$.
It is important to stress that this is not necessarily a clue for physical relation 
with the nearby cloud. The meaning of  these clusters coinciding with the 
theoretically expected $\mathcal{Q} \times D_3$ parameters space
is a possible similarity of the substructures found for the clusters and 
 the clouds that are projected in their direction. 

 The opposite occurs for the clusters that have $\mathcal{Q} > 0.8$ (empty
circles), including NGC~6530 (indicated by error bars), which do not show the same clustering 
structure observed in the  projected clouds.
The values of $\mathcal{Q}$ obtained for these clusters suggest they are more
centrally concentrated, while the nearby clouds  have small scale substructures.
 
In fact, the original geometric structure of the gas distribution is expected no longer be the 
same of the remaining cloud after the cluster formation. It also depends on the initial
conditions of the star-forming environment.
Furthermore, the investigation of clusters evolution 
requires an estimate of their dynamical age, in order to verify if 
their initial structure should have been erased or not (see discussion in Sect.~\ref{sect5}). 

\begin{figure}
\begin{center}
\includegraphics[width=5.5cm,angle=270]{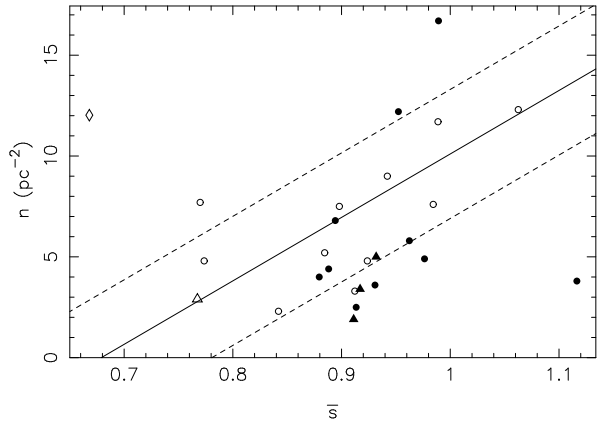}
\caption{Mean surface density compared to $\overline{s}$, which varies more 
with radial clustering. Symbols are the same as Fig.~\ref{fig:ms}. A trend of $\overline{s}$ 
increasing with $n$ is observed only for clusters with $\mathcal{Q} > 0.8$.
}
\label{fig:nxs}
\end{center}
\end{figure}


\subsection{Surface density}

The structural parameters calculated by SG12 were compared to $\mathcal{Q}$,
but no special trend was found for most of them.

We also searched for correlations among structural parameters compared to
the clustering parameters $\overline{m}$ and $\overline{s}$ in separate.
With the exception of NGC~3572  and $\sigma$ Orion, a trend was found for 
clusters having $\mathcal{Q} > 0.8$, which show 
 $\overline{s}$ increasing with average  surface density $n$ (number of stars
per pc$^{2}$). The same does not occur for the other clusters that have substructures,
as seen in the $n \times \overline{s}$ plot of Fig.~\ref{fig:nxs}.
 Also, no trend was found in the comparisons with $\overline{m}$. 
These are expected results, since $\overline{s}$ is more sensitive to radial clustering, 
while $\overline{m}$ varies more  under the presence of substructures \citep{cart04}.

In order to investigate if clustering is related to the cluster concentration
parameter, which is given by tidal radius over core radius, the results
were compared to the relation $\mathcal{Q}$ {\it vs.} $log(R_t/r_c)$ obtained by \citet{sa09}.

According to the definition, $r_c$ is the radius at 
which the surface brightness drops by a factor of two from the central 
value (PZMG10). 

The tidal radius of the sample was estimated by using the relation 
suggested by \citet{saurin}: 
$R_{t} = ({M}/{M_{gal}})^{1/3} d_{GC}$, 
with $M_{gal} = {V^{2}_{GC} d_{GC}}/{G}$, where 
$d_{GC}$ is the distance of the cluster to the Galactic center, 
$V_{GC}$ = 254 km/s and $R_{GC}$ = 8.4 kpc \citep{reid}. The values
of $R_{t}$ are in the range of 7.6 to 13 pc, which give concentration parameter
$log(R_t/r_c)$ that varies from 0.7 to 2, independently of $\mathcal{Q}$.

\citet{sa09} analysed a sample of Galactic open clusters. For the subsample
with $\mathcal{Q} >$ 0.8 they found a correlation 
that is not confirmed by our clusters. 
Contrary to their results, the distribution of all objects in the sample is
not distinguished, having a behaviour similar to the fractal-like subsample analysed by them. 
The same is noted  for $\sigma$~Ori \citep{caballero} that has $R_t$=21 pc, $r_c$=1 pc and $\mathcal{Q}$=0.88. 
These results possibly indicate that our sample is not tidally limited, meaning that tidal
radius is not well defined for them. Yet, $R_t$ estimates depend on many assumptions having large error bars.
For this reason, we preferred not using tidal radius in the present analysis.


\begin{table*}																									
\caption{Results from fractal analysis}																							
\begin{center} 																									
\begin{tabular}{l c c c c c c c c c c}																						
\hline																									
cluster	&	$l^{\rm o}$	&	$\Delta l$	&	$b^{\rm o}$	&	$\Delta b$	&	$N$	&	$\mathcal{Q}$	&	$\overline{m}$	&	$\overline{s}$	&	$D_2$	&	$T_{cr}$ (Myr)	\\
\hline																							
Collinder 205	&	269.21	&	0.12	&	-1.85	&	0.11	&	34	&	0.84$\pm$0.12	&	0.65$\pm$0.21	&	0.77$\pm$0.30	&	1.13$\pm$0.05	&		47.4	\\
Hogg 10 	&	290.8	&	0.05	&	0.08	&	0.05	&	18	&	0.80$\pm$0.12	&	0.79$\pm$0.40	&	0.98$\pm$0.53	&	1.24$\pm$0.09	&		26.5	\\
Hogg 22 	&	338.56	&	0.07	&	-1.14	&	0.06	&	23	&	0.81$\pm$0.11	&	0.76$\pm$0.38	&	0.94$\pm$0.49	&	1.38$\pm$0.03	&		23.1	\\
Lynga 14	&	340.92	&	0.07	&	-1.09	&	0.07	&	15	&	0.75$\pm$0.18	&	0.74$\pm$0.32	&	0.99$\pm$0.53	&	1.40$\pm$0.03	&		15.3	\\
Markarian 38	&	11.98	&	0.06	&	-0.93	&	0.06	&	15	&	0.75$\pm$0.19	&	0.67$\pm$0.22	&	0.89$\pm$0.35	&	1.31$\pm$0.06		&	25.8	\\
NGC 2302	&	219.3	&	0.07	&	-3.12	&	0.07	&	30	&	0.76$\pm$0.15	&	0.67$\pm$0.25	&	0.88$\pm$0.42	&	1.17$\pm$0.05	&		40.2	\\
NGC 2362	&	238.17	&	0.07	&	-5.55	&	0.06	&	22	&	0.81$\pm$0.11	&	0.74$\pm$0.30	&	0.91$\pm$0.37	&	1.34$\pm$0.07	&		21.4	\\
NGC 2367	&	235.59	&	0.07	&	-3.83	&	0.07	&	19	&	0.81$\pm$0.14	&	0.69$\pm$0.25	&	0.84$\pm$0.36	&	1.48$\pm$0.04	&		59.2	\\
NGC 2645	&	264.8	&	0.07	&	-2.9	&	0.08	&	28	&	0.80$\pm$0.12	&	0.72$\pm$0.29	&	0.90$\pm$0.43	&	1.39$\pm$0.04	&		29.4	\\
NGC 2659	&	264.18	&	0.11	&	-1.65	&	0.11	&	55	&	0.83$\pm$0.12	&	0.73$\pm$0.27	&	0.88$\pm$0.40	&	1.39$\pm$0.04	&		45.8	\\
NGC 3572	&	290.71	&	0.07	&	0.2	&	0.07	&	29	&	0.83$\pm$0.13	&	0.64$\pm$0.17	&	0.77$\pm$0.26	&	1.24$\pm$0.10	&		30.6	\\
NGC 3590	&	291.21	&	0.05	&	-0.17	&	0.05	&	13	&	0.84$\pm$0.17	&	0.83$\pm$0.44	&	0.99$\pm$0.45	&	1.24$\pm$0.09	&		20.7	\\
NGC 5606	&	314.85	&	0.06	&	0.99	&	0.07	&	32	&	0.82$\pm$0.14	&	0.75$\pm$0.32	&	0.92$\pm$0.47	&	1.36$\pm$0.04	&		38.6	\\
NGC 6178	&	338.41	&	0.08	&	1.21	&	0.07	&	21	&	0.76$\pm$0.15	&	0.72$\pm$0.29	&	0.95$\pm$0.48	&	1.34$\pm$0.03	&		22.3	\\
NGC 6604	&	18.25	&	0.09	&	1.7	&	0.1	&	33	&	0.74$\pm$0.19	&	0.65$\pm$0.19	&	0.89$\pm$0.40	&	1.42$\pm$0.03	&		37.5	\\
NGC 6613	&	14.18	&	0.11	&	-1.01	&	0.1	&	31	&	0.73$\pm$0.21	&	0.71$\pm$0.28	&	0.98$\pm$0.55	&	1.34$\pm$0.02	&		39.7	\\
Ruprecht 79	&	277.1	&	0.08	&	-0.82	&	0.09	&	53	&	0.77$\pm$0.15	&	0.72$\pm$0.27	&	0.93$\pm$0.45	&	1.51$\pm$0.04	&		53.6	\\
Stock 13	&	290.5	&	0.06	&	1.59	&	0.07	&	9	&	0.71$\pm$0.17	&	0.79$\pm$0.44	&	1.12$\pm$0.63	&	1.46$\pm$0.04	&		38.9	\\
Stock 16	&	306.15	&	0.09	&	0.06	&	0.08	&	19	&	0.76$\pm$0.16	&	0.73$\pm$0.32	&	0.96$\pm$0.52	&	1.41$\pm$0.05	&		43.9	\\
Trumpler 18	&	290.98	&	0.09	&	-0.14	&	0.09	&	86	&	0.73$\pm$0.19	&	0.67$\pm$0.19	&	0.91$\pm$0.40	&	1.24$\pm$0.10	&		65.1	\\
Trumpler 28	&	356.01	&	0.07	&	-0.31	&	0.07	&	17	&	0.81$\pm$0.11	&	0.87$\pm$0.55	&	1.06$\pm$0.69	&	1.31$\pm$0.04	&		18.5	\\
\hline																							
Berkeley 86	&	76.66	&	0.08	&	1.28	&	0.08	&	48	&	0.79$\pm$0.14	&	0.73$\pm$0.30	&	0.92$\pm$0.47	&	1.31$\pm$0.04	&				\\
NGC 2244	&	206.31	&	0.25	&	-2.06	&	0.24	&	256	&	0.74$\pm$0.14	&	0.68$\pm$0.13	&	0.91$\pm$0.27	&	1.39$\pm$0.05	&				\\
NGC 2264	&	202.92	&	0.36	&	2.18	&	0.33	&	225	&	0.71$\pm$0.16	&	0.66$\pm$0.14	&	0.93$\pm$0.30	&	1.49$\pm$0.03	&				\\
NGC 6530	&	6.08	&	0.13	&	-1.32	&	0.12	&	55	&	0.87$\pm$0.10	&	0.66$\pm$0.20	&	0.77$\pm$0.28	&	1.35$\pm$0.03	&				\\
\hline																									
\end{tabular}																									
\label{tab2}																									
\end{center}																									
{\footnotesize																									
Notes: The center of the spatial distribution of the cluster members is indicated in Columns 2-5 showing average galactic coordinates 
and respective $2\sigma$ deviations ($\Delta l$ and $\Delta b$), which roughly indicate the cluster radius.}																									
\end{table*}																									

%

\section{Dynamical Evolution}
\label{sect5}

 No special trend was found when comparing statistical parameters 
with mass and age of the clusters from SG12. This  unexpected
lack of correlation is probably due to the similarities observed in their fundamental 
parameters. 

 Therefore, the analysis of age dependence on cluster 
structure is better illustrated by comparing the sample  with objects spread in a larger 
range of fundamental parameters values, like the young massive clusters (YMC) and stellar 
associations of Milky Way  and other galaxies, using data available in Tables 2 to 4 of 
PZMG10, for instance. Also in this section, the  $\mathcal{Q}$ evolution and the crossing time
as a function of age are compared with observations and theoretical results from literature.

\subsection{Core radius {\it vs} Age }

\begin{figure}
\begin{center}
\includegraphics[width=5.5cm,angle=270]{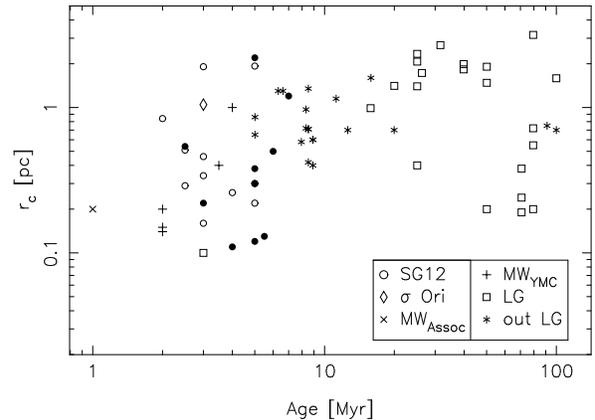}
\caption{Core radius as a function of age of
our clusters (SG12, filled symbols for $\mathcal{Q}<$0.8) and $\sigma$~Ori, 
compared to Associations and Young Massive Clusters of Milky Way (MW),
and clusters of other galaxies of the Local Group (LG) and outside the LG.
}
\label{fig:rcxage}
\end{center}
\end{figure}

It is noted in Fig.~\ref{fig:rcxage} that the sample has age and core radius consistent with 
the Milky Way YMCs, but a large dispersion is observed. 
The inclusion of our objects, as well as $\sigma$ Ori, in this distribution makes 
less clear the expected trend of $r_c$ increasing with age. However, our results 
confirm the prediction by PZMG10 that young clusters with large $r_c$ probably 
were missing in their sample, due to a selection bias. 

It is also interesting to note that our sample tends to fill the upper region in Fig.~\ref{fig:rcxage}
(large $r_c$ values) that is the expected locus of unbound associations, for which there are more available 
data of size than $r_c$ (see Fig.~8 (left) of PZMG10). 

The hypothesis that our clusters probably are expanding objects can be confirmed based on
their dynamical evolution. In the following subsection, the size and mass estimated by 
SG12, were used to investigate how many crossing times the clusters have undergone, which gives an
indication on how dynamically evolved they are.

\subsection{Crossing time}

By using the total mass ($M$) and cluster radius ($R$) from SG12,
the crossing time of the clusters were estimated adopting the expression $T_{cr} = 10({R^3}/{G M})^{1/2}$
from \citet{gieles11}. 
They suggested a boundary distinguishing stellar groups under different dynamical 
conditions that is expressed by the ``dynamical age", also called parameter $\Pi$, 
which is given by the ratio of age and crossing time. Unbound associations (expanding objects) 
have $\Pi <$ 1, while bound star clusters have $\Pi >$ 1.

\begin{figure}
\begin{center}
\includegraphics[width=5.5cm,angle=270]{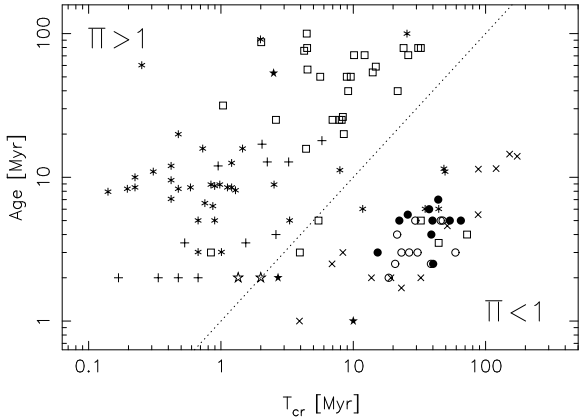}
\caption{Distribution of age {\it vs.} crossing time.
The sample is compared to results from 
\citet{cart04} (\ding{73}) and from PZMG10 (same symbols as Fig.~\ref{fig:rcxage}). 
The groups of Milky Way show two distributions: bound YMCs ($+$) and 
unbound associations ($\times$), separated by a dotted line ($\Pi = 1$).
}
\label{fig:TcrxAge}
\end{center}
\end{figure}

For comparison among different samples, in Fig.~\ref{fig:TcrxAge}  the dynamical 
time ($T_{dyn}$) given by PZMG10 was adopted, assuming that $T_{cr} = 2.8 T_{dyn}$ \citep{gieles11}.
The Milky Way stellar groups are presented in two distinct distributions:
Associations ($\Pi < 1$) and YMCs ($\Pi > 1$). 
In general, the objects in
other Local Group galaxies, like LMC, SMC and M31, as well as 
outside the Local Group, seem to follow the same distribution as YMCs of Milky Way, but are more scattered. 

Figure ~\ref{fig:TcrxAge} clearly shows that all clusters  of the sample are 
more likely associations, following the same trend of Milky Way objects having 
$\Pi < $ 1 that corresponds to unbound expanding groups. No differences
are found when comparing $\Pi$ with the clustering characteristics of our sample, 
meaning that both fractal and smooth radial density profile types of objects  
follow the same age {\it vs.} crossing time correlation.

Among the clusters studied by \citet{cart04}, Taurus ($\Pi$ = 0.1) coincides
with the distribution shown by our sample and other associations, 
while $\rho$ Oph ($\Pi$ = 1.5) and Chamaeleon  ($\Pi$ = 0.74) are closer to IC~348,
which is in the transition region ($\Pi$ = 1). 
IC~2391  ($\Pi$ = 21.2) is found in the region occupied by 
older clusters of other galaxies.
The scattering observed in this region is probably due to observational difficulties
to determine accurate effective radius of distant objects. Our sample is
not affected by this observational constraint.


\subsection{Clustering Evolution}


The values of  $\mathcal{Q}$ and age, observed in the sample, were compared with 
different YSOs classes analysed by \citet{schkl}. We noted 
that $\mathcal{Q}$ values of our clusters are systematically above those obtained with 3D calculations 
($\mathcal{Q} \sim 0.7$) for the YSOs in the same range of ages (2 to 7 Myr). 
\citet[see their Fig. 5]{schkl} suggest that  3D-$\mathcal{Q}$ values tend to be lower than the projected 2D calculations. 
Indeed, a better agreement is found by comparing the results with the calculations for the projections in 2D plans
presented by \citet{schkl}. 


\begin{figure}
\begin{center}
\includegraphics[width=5.5cm,angle=270]{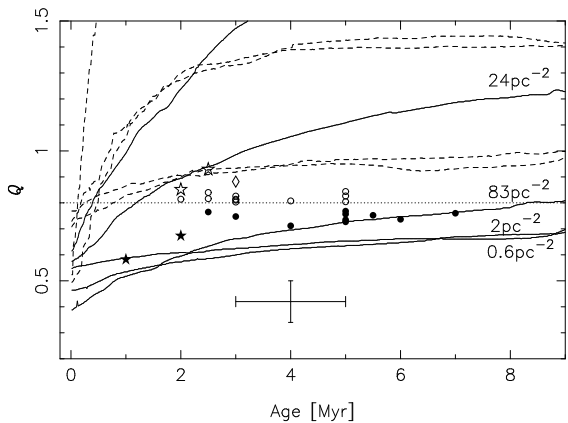}
\caption{Distribution of $\mathcal{Q}$ as a function of age for the sample and other
clusters (same as Fig.~\ref{fig:ms}). Cygnus OB2 is indicated with
error bars. The results are
compared to the simulations with feedback (solid lines)
and without feedback (dashed lines),
indicating the initial densities (number of stars per pc$^2$) for some of the 
models adapted from \citet{pd}.
}
\label{fig:dpQxAge}
\end{center}
\end{figure}

Considering that the evolution of clustering structure may be affected by the initial 
star-forming conditions, in
Fig.~\ref{fig:dpQxAge} the results are compared with the calculations from \citet[see their Fig. 3]{pd} that
used models {\it with-} and {\it without feedback} to simulate the effects of the presence (or not) of 
ionizing sources affecting the star formation \citep{dale12,dale13}. 
Even taking in account the large error bars on $\mathcal{Q}$, as illustrated in Fig.~\ref{fig:QxDf}, the 
distribution of $\mathcal{Q}$ values in the sample, as a function of age, is consistent 
with {\it feedback models}, in particular those assuming  
low initial densities ($\Sigma_1$ = 0.6 to 83 stars pc$^{-2}$). 
\citet{pd} suggest that the lowest density models are the mostly affected by feedback, which are the only ones retaining some 
substructure for 5 to 10 Myr. Since clusters with low densities have longer relaxation times, 
their substructures are not erased due to negligible stellar mixing.

Other young clusters from literature were also included in Fig.~\ref{fig:dpQxAge}. 
Excepting Cygnus OB2 \citep{wright}, 
the other clusters like $\rho$ Oph, Taurus, Chamaeleon, IC348 \citep{schkl}, and $\sigma$ Ori \citep{caballero}
are similar to our sample, coinciding with  low density {\it feedback models} from \citet{dale12,dale13}.

The results are also comparable to those from \citet{pd14} for the 
evolution of $\mathcal{Q}$ based on simulations of different values of virial ratio $\alpha_{vir}$,
the total kinetic energy over the potential energy. Our clusters have a 
distribution very similar to the simulation of $\alpha_{vir}$ = 1.5 (see their Fig.~3g,h,i),
which represents a globally supervirial fractal that is ``hot" and unbound, according the
terminology introduced by them.
The same is found for Cygnus OB2, a very substructured 
cluster ($\mathcal{Q}=0.4$), which is better represented by
calculations for unbound regions under supervirial conditions
\citep{wright}.

\section{Conclusions}
\label{sect6}

In the present work, the methodology used by FGH12 was adopted to perform a fractal analysis of 
the sample studied by SG12,  aiming to enlarge our previous study  
for the clusters NCG~6530, Berkeley~86, NGC~2244, and NGC~2264. About half of the
sample (52\%) shows substructures, indicated by the measured parameter $\mathcal{Q} < 0.8$,
while the other 48\% of the objects are centrally concentrated ($\mathcal{Q} > 0.8$).
According to \citet{pd14}, regions having low $\mathcal{Q}$ ($\mathcal{Q} <0.8$ or 1) must be
dynamically young.

The projected spatial distribution of very likely members, which have membership probability 
$P > 50\%$, shows a $\overline{m}$ {\it vs.} $\overline{s}$ correlation  that is confirmed 
by results from literature for other young clusters. On the other hand, the
sub-sample of candidate members ($P < 50\%$) do not follow this trend.

Centrally concentrated clusters of the sample have 
$\overline{s}$ increasing with the mean spatial density ({\it n}), which is an expected result since
$\overline{s}$ shows more variation for radial clustering. This could also be
the explanation for the lack of trend in the {\it n vs.} $\overline{s}$ for objects with fractal
clustering.  However, $\sigma$ Ori shows a too high surface density, being out of the correlation with 
$\overline{s}$ found by us. This discrepancy is possibly due to uncertainties on the parameters adopted for $\sigma$ Ori
($N_T$ = 340 or $R$ = 3 pc).

Visual extinction maps of clouds nearby the clusters were used in order
to study the geometric structure of gas and dust distribution, by measuring their fractal dimension $D_2$. 
A transform function was adopted to compare the results, measured on projected 2D maps,
with simulations based on 3D models. 
Intermediate values $2.1 < D_3 < 2.5$ were found, which correspond more likely to 
substructures than to smooth radial profiles.

The distribution of fractal dimension $D_3$ (estimated for the observed clouds) as a function 
of $\mathcal{Q}$ parameter (measured for the clusters) shows that half of the sample 
tends to follow a  theoretically expected $\mathcal{Q}$ $\times ~D_3$ relation, calculated for 
artificial data \citep{cart04,sa09}. 
These clusters have
$\mathcal{Q}<0.78$ that indicates the presence of substructures, similar to those observed in the nearby clouds. 
On the other hand, 12 clusters (48\%), among them NGC~6530, have $\mathcal{Q} > 0.8$, which corresponds 
to radial distribution of stars that does not coincide with the structure of the clouds.

It should be expected that these two groups of clusters
could have had different virial conditions in the early formation, which are related to different
thermal paths: cold collapse for objects centrally concentrated, or warm collapse for those
showing substructures \citep{delgado13}. Therefore, differences on dynamical evolution could also be expected,
but they were not observed for these two subsamples. 

The analysis of dependence of age on parameters 
like core radius and crossing time show that our clusters have many similarities among
each other. They are also similar to other young stellar groups of Milky Way, in particular the
unbound associations, as indicated by the dynamical ages $\Pi < 1$ found for the entire sample. 



By comparing the evolution of $\mathcal{Q}$ with theoretical studies, the clusters
show a distribution similar to simulations  that include feedback of ionizing source in 
the  models adopted by \citet{pd}. According these authors, the calculations using low values of density are the mostly affected by 
feedback, corresponding to the only simulations retaining some substructure for 5 to 10 Myr, which 
appears to be the case of our sample.
Other young clusters from literature also coincide with these models. A slight
tendency is observed for objects that have smooth density profile ($\mathcal{Q}>0.8$), 
like IC~348 and $\rho$ Oph, being more similar to the calculation with $\Sigma_i$ = 24 stars pc$^{-2}$. 
On the other side, Taurus, which shows fractal structure, coincides with 
the $\Sigma_i$ = 0.6 stars pc$^{-2}$ model.  Our objects are found in between this range
of densities.

The observed distribution of $\mathcal{Q}$ as a function of age  was also compared to
the simulations by \citet{pd14} assuming different values of virial ratio ($\alpha_{vir}$). The results 
are quite similar to the calculations that use $\alpha_{vir}=1.5$, corresponding to ``supervirial" fractal.

Since no difference concerning the $\mathcal{Q}$ evolution is noted for our objects, independently if
they are fractal or have radial density profile, it is not evident to state if they had or not 
similar conditions in the early formation.
 
Additional discussions on the cluster formation conditions must include the initial density distribution 
of parental clouds, as well as the effects of turbulence. According \citet{girichidis}, compressive modes 
in a flat density profiles tend to form substructures, instead of centrally concentrated distributions 
of stars. The fractal dimension calculated for clouds, projected in the direction 
of the studied clusters, allow us
to infer the geometric distribution of the dense regions. A more detailed study on dynamical conditions 
of the gas is required to better understand the scenario of clusters formation and their relation with their 
parental clouds.

\section*{Acknowledgements}

We thank the anonymous referee for significant comments and useful suggestions.
Part of this work was supported by CAPES/Cofecub Project 712/2011, 
FAPESP Proc. No. 2010/50930-6, and
CNPq Projects: 142849/2010-3 (BF) and 142851/2010-8 (TSS).
This publication makes use of data products from the Two Micron All Sky Survey, 
which is a joint project of the University of Massachusetts and the 
Infrared Processing and Analysis Center/California 
Institute of Technology, funded by the National Aeronautics and 
Space Administration and the National Science Foundation.

\label{lastpage}
\end{document}